# Measuring the non-Abelian Quantum Phase with the Algorithm of Quantum Phase Estimation

Seng Ghee Tan (1)[†], Son-Hsien Chen (2), Ying-Cheng Yang (3), Yen-Fu Chen (1),
Yen-Lin Chen (2), Chia-Hsiu Hsieh (2)

(1) Department of Optoelectric Physics, Chinese Culture University, 55 Hwa-Kan Road,
Yang-Ming-Shan, Taipei 11114, Taiwan
(2) Department of Applied Physics and Chemistry, University of Taipei, Taipei 10048, Taiwan
(3) Department of Physics, National Taiwan University, Taipei 10617, Taiwan

**ABSTRACTS**

We propose an approach to measure the quantum phase of an electron in a non-Abelian system using the algorithm of Quantum Phase Estimation (QPE). The discrete-path systems were previously studied in the context of square or rectangular rings. Present focus is on measuring the quantum phases. The merit of the algorithm approach is two-fold. First off, it eliminates the need for an interferometric set up. Quantum phase is measured by reading off of measurable qubit states of the QPE modules. Secondly, the QPE works by subjecting the quantum state to a sequence of quantum computing operations that eventually map the phase information into measurable qubit states. All the operations are realizable by standard quantum computer gates and algorithms, placing the new effort within the reach of standard quantum computational framework.

Corresponding author:
(†) Seng Ghee Tan (Prof)
Department of Optoelectric Physics,
Chinese Culture University,
55, Hwa-Kang Road, Yang-Ming-Shan, Taipei, Taiwan 11114 ROC
(Tel: 886-02-2861-0511, DID:25221)

** Email: csy16@ulive.pccu.edu.tw; tansengghee@gmail.com
PACS:

## INTRODUCTION

The concept of quantum phase has added new layers of mystery to the behavior of quantum particles, particularly in the context of geometric and topological phenomena. A quantum phase [1,2] can generally be decomposed into two components: the geometric phase and the dynamic phase. The geometric phase [3-5] arises from the intrinsic geometry of the quantum state's evolution in Hilbert space. This phase is independent of the system's energy and depends solely on the path traversed by the quantum state. In contrast, the dynamic phase is directly related to the time evolution of the system's Hamiltonian and accumulates proportionally to the system's energy over time. In an electronic transport system where carrier spin precesses about the momentum-dependent magnetic fields, a non-Abelian [6-8] type of total quantum phase that comprises the geometric and the dynamic components arises. Consider two electronic particles traveling in specific paths, quantum entanglement is found to have an intricate effect [9, 10] on the quantum phases of the bipartite system. In fact, the geometric and dynamic components of the quantum phase can be distilled by tuning the entanglement strength. The reverse implication of the above is thus far-reaching. Quantum phases, when properly measured, can be used to understand and determine the strength of quantum correlation – with entanglement being the most prospective candidate.

Historically, the measurement of phases has been a challenging task, primarily relying on the interference-based techniques. The Mach-Zehnder [11-13] is a space-based interferometry that typically involves constructing interferometers that beam-splits and reconvenes the wave - light or matter with adaptation - to extract the phase information accumulated over the physical routes travelled. Phase differences typically arise from refractive index changes, gravitational effects, magnetic field effects, or simply path length differences along the spatial routes. The Ramsey [14-16] is, on the other hand, a time-based interferometry that relies on quantum superposition and the interference of the superposition states within the same spatial location. Quantum phase difference arises from the Hamiltonian evolution of the internal quantum states in time (e.g. spin or energy levels due to the effects of magnetic fields), or geometric phases due to parameter variations.

In this paper, we propose an approach to measure the non-cyclic quantum phase in a spin-orbit system using the algorithm of Quantum Phase Estimation (QPE). Such systems were previously studied in the context of square or rectangular rings [9, 10]. Our focus in this paper is on measuring the quantum phases, and the rectangular structure merely plays the role of a convenient non-Abelian phase generator. QPE is a cornerstone algorithm in quantum computing [17, 18], originally developed for eigenvalue estimation in unitary operators. It leverages on the principles of quantum Fourier transform and controlled operations to extract phase information encoded in the eigenstates of a given operator. The merit of the algorithm approach is two-fold. First off, it eliminates the need to build conventional interferometric apparatus, which demands precise beam alignment in Mach-Zehnder, and accurate application of temporal pulses in Ramsey. In the algorithm approach, electron carrying a quantum phase is fed directly into the QPE modules. As phase information is encoded in the probability amplitudes of the quantum state, measurement is carried out by simply reading off of the qubit states on the QPE output. In other words, phase information has thus been decoded. Secondly, QPE operates directly within the quantum computational framework. At its core, QPE works by applying a sequence of controlled-unitary operations to the state that carries an evolutionary phase, and performing an inverse quantum-Fourier-transform (IQFT) to map the phase information into measurable qubit states. All these operations, that include state preparation and evolution, controlled-unitary and IQFT operations, are implementable using standard quantum computer gates and algorithms (e.g. Hadamard, CNOT, phase gates, and the IQFT algorithm). Our approach would thus be compatible with the architectures of quantum computers. The full system would thus comprise a spin-orbit, quantum-phase generator and a QPE measurement device, as shown in Figure 1. The phase generator comprises the quantum dot that controls the emission and collection of electron, and the spin-orbit paths that generates the quantum phase.

**Quantum Phase Generator**

The phase generator paths are made of spin-orbit materials, and shaped to a trajectory of discrete steps. A neat example would be an L-shaped structure with paths denoted by numerals 1 and 2 as shown in Figure 1. Travel length of path 1 corresponds to precession angle $\eta$, and electron evolution here is modeled by the quantum rotation gate $R_x$ with a negative angle. Likewise for path 2 where travel length corresponds to $\delta$, electron evolution here is modeled by the quantum rotation gate $R_y$ with a positive angle. Operation wise, electron is initially confined in the emitter quantum dot before being gated for release into path 1. The collector quantum dot would collect electron for measurement with the QPE later. The QPE modules are integrated into the full system via the quantum dots as shown in Figure 2.

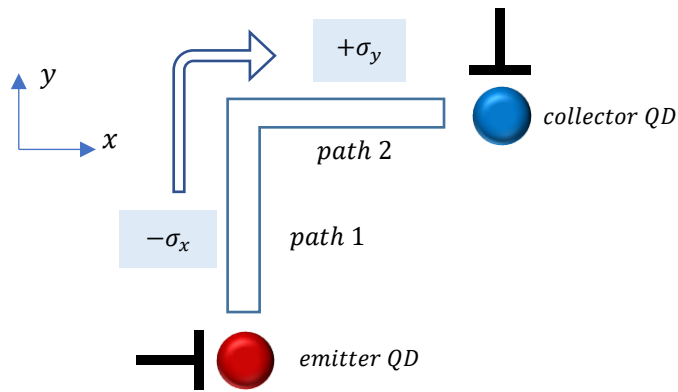

Figure 1. An L-shaped structure (spin-orbit material) comprising of path-1 and path-2 act as a quantum phase generator. Electron is launched into path-1 via the emitter quantum dot and is collected at the collector quantum dot for measurement.

As electron traverses through the discrete steps of the L structure, non-Abelian phase is generated via spin-orbit coupling. The spin-orbit physics that governs electron dynamic is described in a linear form as follows

$$H = \alpha(k_x\,\sigma_y - k_y\,\sigma_x) + \frac{\hbar^2\,k^2}{2m}$$

(1)

One example of the Hamiltonian above is a 2D Rashba spin-orbit coupling system where $\alpha$ is the Rashba constant. Other useful non-Abelian [7, 19-22] platforms may be designed from Weyl, topological or numerous other spintronic surfaces. Considering a quantum state $\psi$ evolving over time in Hilbert space from 0 to $\tau$, the geometric phase is expressed as follows

$$\gamma = \arg\langle\psi(0)|\psi(\tau)\rangle + i\int_0^\tau \langle\psi(t)|\dot\psi(t)\rangle dt$$

(2)

The first and second terms on the right-hand side represent the total quantum phase and the dynamic phases, respectively, where $\arg(x + iy) \equiv \tan^{-1}\left(\frac{y}{x}\right)$. In the Schrodinger picture, the electron evolves as $|\psi(t)\rangle = U|\psi(0)\rangle$, where $U = e^{-i\frac{H}{\hbar}t}$ is unitary due to the Hermitian Hamiltonian. Conservation of norm and reversibility is secured in the evolution. The spin-orbit Hamiltonian can be thought of as the Zeeman energy arising due to the effective magnetic fields. The magnetic fields are momentum dependent, i.e. $(B_x, B_y) \propto (-k_y, k_x)$. Referring to the discrete trajectory of Figure 1, the unitary evolution operator for path-1 and path-2 is

$$U_{12} = U_2.U_1 = e^{-\frac{i\sigma_y}{2}\delta}.e^{-\left(\frac{i(-\sigma_x)}{2}\right)(\eta)}$$

(3)

The evolution angles are: $\frac{\eta}{2} = \left(-\alpha\frac{k_x f t}{\hbar}\right) = \left(-\frac{\omega t}{2}\right)$, and $\frac{\delta}{2} = \left(\alpha\frac{k_y f t}{\hbar}\right) = \left(\frac{\omega t}{2}\right)$, where $\frac{\omega}{2} = \alpha\frac{k_{(x,y)f}}{\hbar}$ represents the Larmor frequency proportional to the strength of the effective magnetic field. Evolution angles $\frac{\eta}{2}$ and $\frac{\delta}{2}$ correspond to the phase distances travel by the electrons. The evolution operator in matrix form is $e^{-i(\boldsymbol{\sigma}.\boldsymbol{n})\frac{\theta}{2}} = \cos\frac{\theta}{2} - i\boldsymbol{\sigma}.\boldsymbol{n}\sin\frac{\theta}{2}$. To be compatible with quantum computers, we express the $U$ operator in terms of standard quantum rotation gates $U = R_y(\delta)R_x(-\eta)$. Mathematically, operators $R_{x(y,z)}$ are spin rotation matrices about effective field directions along $x(y,z)$, respectively.

**QPE Measurements**

To prepare for QPE measurement, electron is required to travel through the discrete step connecting path-1 and path-2. Along the way, electron would be subject to two QPE measurements. Initial quantum state $|0\rangle = \binom{1}{0}$ is prepared in the emitter quantum dot on the bottom-left end. Considering that $U_1 = e^{+i\sigma_x\frac{\eta}{2}}$ for path-1, initial state is expressed in the basis $x$ as follows:

$$|\psi_0\rangle = |0\rangle = \frac{1}{\sqrt{2}}|+x\rangle + \frac{1}{\sqrt{2}}|-x\rangle$$

(4)

As the electron traverses a phase distance corresponding to $\frac{\eta}{2}$ along path-1, the individual eigen-basis evolves as

$$|\psi_1\rangle = \frac{1}{\sqrt{2}}\left(e^{+\frac{i\eta}{2}}|+x\rangle + e^{-\frac{i\eta}{2}}|-x\rangle\right)$$

(5)

As the electron takes to path-2 where $U_2 = e^{-i\sigma_y\frac{\delta}{2}}$, it would be helpful to now view the state in the $y$ basis. The choice of the $y$ basis simply means that the quantum state would be composed of the eigen-basis of the Hamiltonian. Upon traversing phase distance $\delta$, the quantum state is:

$$|\psi_2\rangle = \cos\left(\frac{\pi}{4}-\frac{\eta}{2}\right)e^{-\frac{i\delta}{2}}|+y\rangle + \sin\left(\frac{\pi}{4}-\frac{\eta}{2}\right)e^{+\frac{i\delta}{2}}|-y\rangle$$

(6)

Phase information of path-1 is now fully captured in the probability amplitudes of quantum state $|\psi_2\rangle$. We would now take a pause from proceeding to measurement. Instead, we shall first focus on phase due to path-1, which, when viewed in the $y$ basis, is encoded in the amplitudes of $\cos\left(\frac{\pi}{4}-\frac{\eta}{2}\right)$ and $\sin\left(\frac{\pi}{4}-\frac{\eta}{2}\right)$. An QPE module, known henceforth as the QPEV, is incorporated into the semi-rectangular ring at the end part of path-1 as shown in Figure 2.

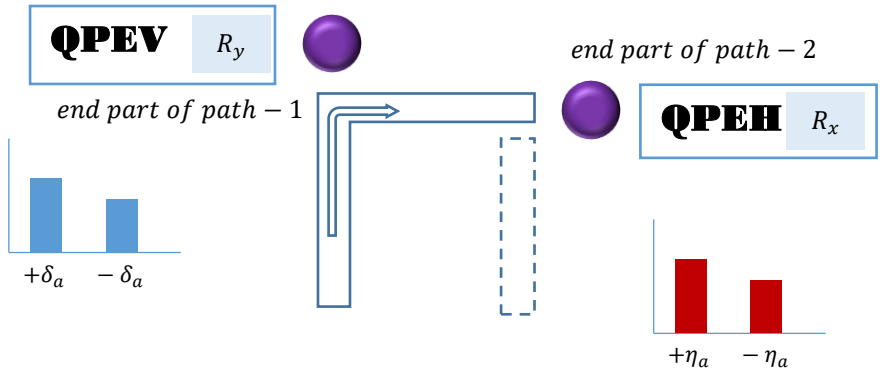

Figure 2. Electron is collected at the end part of path-1 for measurement using the QPEV module. Likewise, electron is collected at the end part of path-2 for measurement with the QPEH module.

Quantum state $|\psi_1\rangle$ is fed into the QPEV through the quantum dot. Rotational gates $R_y$ within the QPEV are manually set to rotate spin by an auxiliary angle of $\delta_a$, achieved largely by setting $t$ to a specific value. The IQFT would produce, upon multiple measurements, a statistical distribution of eigenvectors corresponding to phase $\pm\frac{\delta_a}{2}$. This is because with operator $R_y$, state $|\psi_1\rangle$ will be measured in the projected basis of $y$. And with specific $t$ values, phase $\pm\frac{\delta_a}{2}$ corresponding to $|\mp y\rangle$ are extracted by the QPEV. From the expression of $|\psi_2\rangle$, one can imagine as though the electron has traveled a phase distance of $\delta_a$ within the QPEV, generating a quantum state of

$$|\psi_{1e}\rangle = \cos\left(\frac{\pi}{4} - \frac{\eta}{2}\right) e^{-\frac{i\delta_a}{2}}|+y\,\rangle + \sin\left(\frac{\pi}{4} - \frac{\eta}{2}\right) e^{+\frac{i\delta_a}{2}}|-y\,\rangle$$

$$(7)$$

Of course, $\pm\frac{\delta_a}{2}$ have nothing to do with the actual evolutionary phase distance. They are simply the phase information encoded in the eigenvectors of $R_y$. With $t$ properly set, eigenvalues can be retrieved from reading off of the IQFT. It is worth noting that while the eigenvalues of the projected basis locate the amplitude to be measured at $\pm\frac{\delta_a}{2}$, they have no bearing to the actual amplitude values. One might wonder whether the QPEV can be designed with gates $R_x$ (instead of $R_y$) to measure $e^{\pm\frac{i\eta}{2}}$ of $|\pm x\,\rangle$ straight out of $|\psi_1\rangle$. The answer is negative as the evolutionary $e^{\pm\frac{i\eta}{2}}$ are ready phase factors of $|\pm x\,\rangle$ that would be transparent to the algorithm of QPE. The QPEV would simply measure the phases encoded in the eigenvectors of $R_x$. These phases of $|\pm x\,\rangle$ are irrelevant to the evolutionary phases accumulated by electron traversing the phase generator.

In the following, a detailed description of the QPEV designed specifically to measure the evolutionary phase of path-1 is now provided. Figure 3 shows a schematic quantum circuit layout of the QPEV, as well as the phase generator for path-1. The spin-orbit phase generator is modelled by the quantum gate $R_x(-\eta)$. On the other hand, QPEV comprises the rotational gates of $R_y(\delta_a)$, and quantum phase is read off of the IQFT.

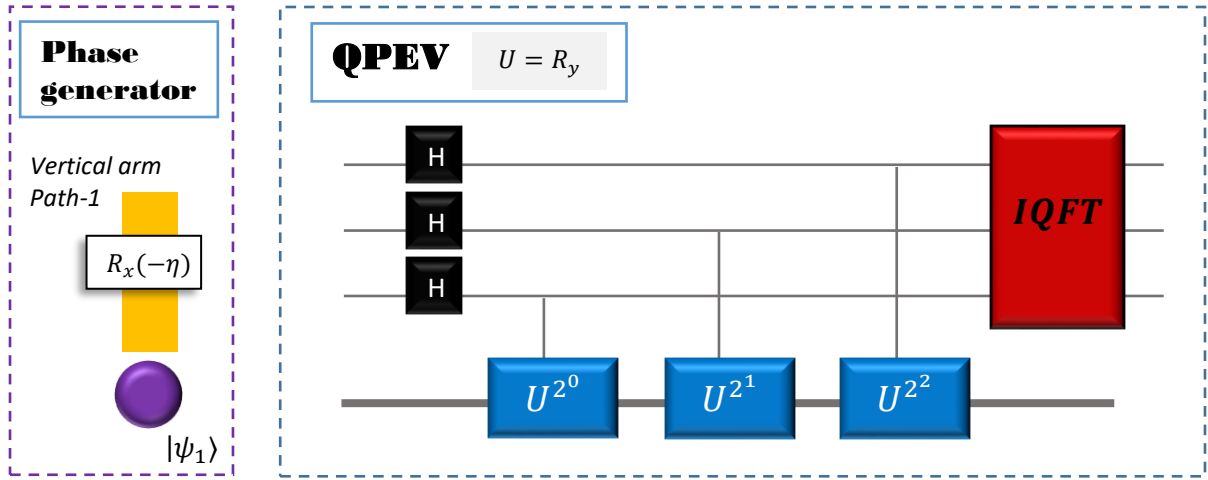

*FIG.3. Left: Model of the spin-orbit phase generator for path-1. Right: Model of the QPEV that measures the amplitude of electron encoded by $\eta$.*

To lend further credence to the method discussed above, we will carry out quantum computation using the IBM Qiskit. We will first make a theoretical estimate of the expected outputs from the QPEV measurement. Let us choose a $t$ value such that $\delta_a = \frac{\pi}{4}$ is fixed for the internal $R_y$ gates. Referring to $|\psi_{1e}\rangle$ of Eq.(7) and Table 1 for the binary representation of $\delta_a$, one would expect the IQFT readout to comprise histogram bars at 0.00010 and 0.11110. Let us now denote the amplitudes as follows: $C = \cos\left(\frac{\pi}{4} - \frac{\eta}{2}\right)$ and $S = \sin\left(\frac{\pi}{4} - \frac{\eta}{2}\right)$. We will now let electron travel path-1 for a phase distance of $\eta = +\frac{\pi}{3}$. In spin-orbit physics, spin rotates about an effective $-B_x$ field by $+\frac{\pi}{3}$ according to $e^{\frac{-i(-\sigma_x)}{2}\left(+\frac{\pi}{3}\right)}$. At the quantum computer level, this is to be modelled by applying gate $R_x\left(\eta = -\frac{\pi}{3}\right)$ to input state $|0\rangle$, which would represent spin rotating about a $+B_x$ field by $-\frac{\pi}{3}$. The two equivalent pictures would give an estimate of $C^2 = 0.933$ and $S^2 = 0.066$ as shown in Table 1.

Table 1. Binary representation of $\delta_a$ values and the corresponding amplitudes for $R_x\left(-\frac{\pi}{3}\right)$.

| Binary readout on the IQFT | Quantum gate: $\delta_a = \frac{\pi}{4} \rightarrow \left(e^{-\frac{i\pi}{8}}, e^{+\frac{i\pi}{8}}\right)$ | Probability Amplitudes |
|---|---|---|
| 0.00010 | binary interpretation:<br>$\left(+\frac{\delta_a}{2}\right) = \frac{1}{16}(2\pi) = +\frac{\pi}{8}$ | $S^2 = 0.066$ |
| 0.11110 | binary interpretation:<br>$\left(-\frac{\delta_a}{2}\right) = \frac{15}{16}(2\pi) = -\frac{1}{16}(2\pi) = -\frac{\pi}{8}$ | $C^2 = 0.933$ |

Output of the IBM Qiskit shows that the quantum computation results are very close to the theoretical estimates above. With 10 Hadamard gates, and 10,000 shots in the number of measurement, Figure 4 delivers $|C|^2 \sim 0.9319$ and $|S|^2 \sim 0.0681$, matching the theoretical estimates fairly accurately.

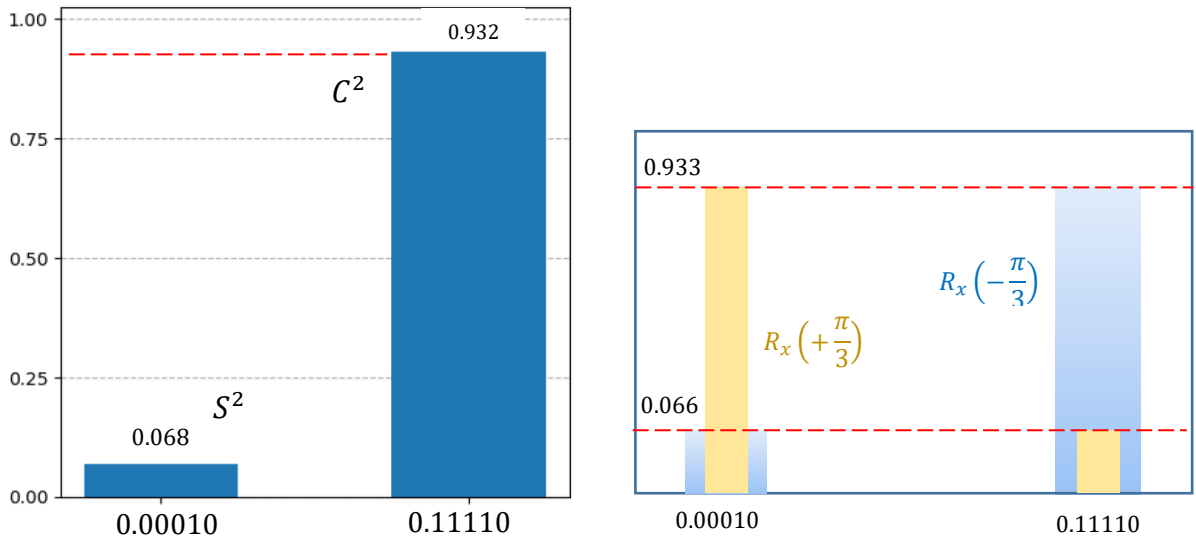

FIG.4. Qiskit modeling of the QPEV with $R_x\left(\eta = -\frac{\pi}{3}\right)$ to measure the quantum phase of electron on completion of path-1. Histogram bars of $0.0001$ and $0.1111$ correspond to $S^2$ and $C^2$, respectively.

As a matter of fact, $e^{\pm\frac{i\eta}{2}}$ are phase factors of eigenvectors $|\pm x\rangle$. But quantum phase for electron traversing path-1 is a perennial zero as follows:

$$\langle 0|U_1|0\rangle = \cos\frac{\eta}{2} \;\rightarrow\; \arg\langle 0|U_1|0\rangle = 0$$

(8)

Therefore, regardless of the quantum computation outcome for $C^2$ or $S^2$, phase is zero in the case of path-1, which is Abelian. However, quantum computation outcome will be important for the non-Abelian phase when electron takes to path-2, on completion of path-1. The results of Figure 4 are also important to confirm that the QPE method works correctly for the amplitudes.

We have described the concept and the apparatus to measure the evolutionary phase due to Abelian path-1. We demonstrated with IBM Qiskit that quantum computation results are in agreement with theoretical estimates for the amplitudes. We have also expounded the particular need to use controlled gates $R_y$ for the projected basis of a phase evolution process modeled by $R_x$ as clearly seen in path-1. With this, we will now proceed to measure the non-Abelian phase of electron traversing the vertical path-1 and the horizontal path-2, which prescribe a quantum state of $|\psi_2\rangle$ as shown in Eq.(6). This is a result of $U = R_y(\delta)R_x(-\eta)$ where phase evolution along path-2 is modelled by $R_y(\delta)$. Following the earlier approach of projected measurement, controlled gates $R_x$ are deployed to measure the non-Abelian phase. An additional QPE module, known henceforth as the QPEH, is incorporated into the semi-rectangular ring on the end part of path-2 as shown in Figure 2. Quantum state of Eq.6 is re-expressed in basis $x$ as follows

$$|\psi_2\rangle = \frac{1}{\sqrt{2}}(A|+x\rangle + B|-x\rangle)$$

(9)

where

$$A = (C+S)\cos\left(\frac{\pi}{4} - \frac{\delta}{2}\right) + i(C-S)\sin\left(\frac{\pi}{4} - \frac{\delta}{2}\right)$$

(10)

$$B = (C+S)\sin\left(\frac{\pi}{4} - \frac{\delta}{2}\right) - i(C-S)\cos\left(\frac{\pi}{4} - \frac{\delta}{2}\right)$$

(11)

As in the case of QPEV, state $|\psi_2\rangle$ will be fed into the QPEH via a quantum dot. Rotational gates $R_x(+\eta_a)$ within the QPEH are manually set to rotate spin by an auxiliary angle, once again achieved largely by setting $t$ to a specific value. The IQFT would produce, upon multiple measurements, a statistical distribution of eigenvectors corresponding to phase $\pm\frac{\eta_a}{2}$. One can imagine as though the electron has traveled a phase distance of $\eta_a$ within the QPEH, generating a quantum state of

$$|\psi_{2e}\rangle = e^{-\frac{i\sigma_x}{2}\eta_a}|\psi_2\rangle = \frac{1}{\sqrt{2}}\left(A|+x\rangle\, e^{-i\frac{\eta_a}{2}} + B|-x\rangle\, e^{+i\frac{\eta_a}{2}}\right)$$

(12)

As explained earlier, $\pm\frac{\eta_a}{2}$ are, this time, simply the phase information encoded in the eigenvectors of $R_x$. With $t$ properly set, $\pm\frac{\eta_a}{2}$ provide the locations where amplitudes of $|A|$ and $|B|$ are measured. Once again, $\pm\frac{\eta_a}{2}$ have no bearing to the amplitudes. Figure 5 provides a detailed description of the QPEH, in a schematic circuit lay-out designed specifically to measure the evolutionary phase due to electron traversing path-1 and path-2. As shown on the left, quantum gate $R_x(-\eta)$ is used to model the actual phase dynamic of electron traversing path-1. On the other hand, QPEV comprises the rotational gates of $R_y(+\delta_a)$, and quantum phase is read off of the IQFT. At this point, it is not hard to recognize that two QPE measurements are required to complete the measurement of the non-Abelian quantum phase. QPEV was designed to measure the electron phase upon its completion of path-1. QPEH would complete the task by measuring phase upon completion of path-1 and path-2. We will carry out quantum computation using the IBM Qiskit.

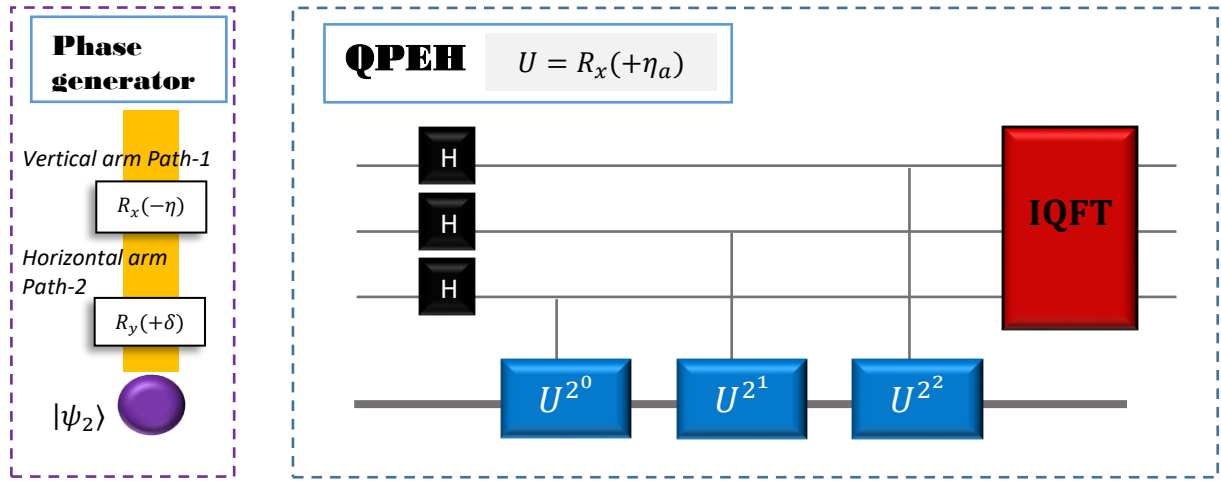

FIG.5. Left: Model of the spin-orbit phase generator for path-1 and path-2. Right: Model of the QPEH circuit that measures the amplitude of electron encoded by $\eta$ and $\delta$.

We will first make a theoretical estimate of the expected outputs from the QPEH measurement. Recalling Eqs.(10) and (11), one has

$$A = \sqrt{(1 + 2SC\sin\delta)}\ e^{i\gamma_1}\ ;\quad B = \sqrt{(1 - 2SC\sin\delta)}\ e^{i\gamma_2}$$

(13)

where $e^{i\gamma_1}$ and $e^{i\gamma_2}$ would simply be the ready phase factors transparent to the QPE algorithm, in this case QPEH. Explicitly, Eq.(12) can be written as

$$|\psi_{2e}\rangle = \frac{1}{\sqrt{2}}\left(|A|\ e^{i\gamma_1 - i\frac{\eta_a}{2}}\ |{+}x\rangle\ +\ |B|\ e^{i\gamma_2 + i\frac{\eta_a}{2}}\ |{-}x\rangle\right)$$

(14)

Let us now choose a $t$ value such that $\eta_a = \frac{\pi}{4}$ is fixed for the internal $R_x$ gates. The QPEH operation is $R_x\left(\eta_a = \frac{\pi}{4}\right)$. By contrast, the QPEV operation was $R_y\left(\delta_a = \frac{\pi}{4}\right)$. Referring to $|\psi_{2e}\rangle$ and Table 1 for the binary representations, one would expect the IQFT readout to comprise histogram bars corresponding to $\frac{1}{2}|A|^2$ and $\frac{1}{2}|B|^2$ at, respectively, 0.11110 and 0.00010. We had earlier on measured $C$ and $S$ using the QPEV with an auxiliary parameter $\delta_a$. We will now let electron travel path-1 for a phase distance of $\eta = \frac{\pi}{3}$, and path-2 for phase distance $\delta = \frac{\pi}{3}$, which can be modelled by applying $R_y\left(\delta = \frac{\pi}{3}\right) R_x\left(\eta = -\frac{\pi}{3}\right)$ to input state $|0\rangle$. This would give an estimate of $\frac{1}{2}|A|^2 = 0.7165$ and $\frac{1}{2}|B|^2 = 0.2835$, for the respective bars. To measure $|A|$ and $|B|$ though, quantum computation using the IBM Qiskit is carried out with auxiliary $\eta_a = \frac{\pi}{4}$, in the same manner that we had earlier on measured $C$ and $S$ using the QPEV with auxiliary $\delta_a = \frac{\pi}{4}$. With the 10 Hadamard gates, and 10,000 shots in the number of measurement, Figure 6 delivers results of $\frac{1}{2}|A|^2 \sim 0.7146$ and $\frac{1}{2}|B|^2 = 0.2854$.

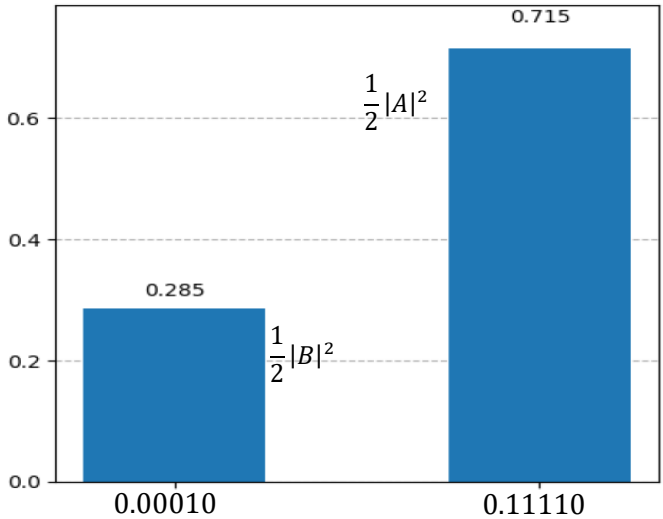

*FIG.6. Qiskit simulation of the QPEH to measure the non-Abelian phase of an electron traversing path-1 and path-2 of the spin-orbit phase generator.*

Having affirmed the correctness of the method above in determining the amplitudes, we will turn to the discussion of how the approach is used to measure the overall, non-cyclic quantum phase of an electron traversing a non-Abelian path, hereinafter known as $\theta$. The non-Abelian quantum phase factor is:

$$\langle 0|\psi_2\rangle = \frac{1}{2}(A + B) = \sqrt{\left(\frac{1}{2} + SC\cos\delta\right)}\, e^{i\theta}$$

(15)

This would lead to the non-Abelian phase of

$$\theta = \tan^{-1}\left(\frac{S-C}{S+C}\tan\frac{\delta}{2}\right)$$

(16)

As discussed, amplitudes of $S, C$ were measured by the QPEV module. Meanwhile, $|A| = \sqrt{(1 + 2SC\sin\delta)}$ means that $\delta$ can be inferred from all the measured amplitudes as follows:

$$\sin\delta = \frac{|A|^2 - 1}{2SC}$$

(17)

Finally, the non-Abelian phase $\theta$ is calculated using Eq.(16).

**CONCLUSION**

We have proposed an algorithm-based approach of QPE to measure the non-Abelian quantum phase of electrons traversing a discrete path. An L-shaped spin-orbit system is used to generate the non-Abelian phase as electron traverses the structure. Two specific schemes, namely the QPEV and the QPEH are designed to measure the amplitudes. When cast in the projected measurement basis, state amplitudes are encoded by the evolutionary phase information. Electron evolution along paths 1 and 2 are modeled with the standard quantum rotational gates. Measured amplitudes via the quantum computation approach of IBM Qiskit were consistent back-of-the-envelope theoretical estimates, affirming the correctness of the quantum approach. We have therefore, proposed a QPE-based approach, with accuracy confirmed, that the non-Abelian phase of $\theta$ can be determined via direct extraction of the electron phase information via its amplitudes. Finally, our work establishes a direct interface with the quantum computational techniques, by translating phase measurement into a QPE-compatible protocol. Our approach simplifies phase measurement, and unlocks new possibilities for studying quantum phases in scalable, gate-based quantum processors, possibly enabling future quantum metrology without external interferometric instability.

## Acknowledgement

We would like to thank the National Science and Technology Council of Taiwan for supporting this work under Grant. No.: NSTC 113-2112-M-034-002 and NSTC 114-2112-M-034-001.

## Data Availability Statement

The data that support the findings of this study are available from the corresponding author upon reasonable request.